\documentclass[manuscript,screen,nonacm]{acmart}

\usepackage{tabularx}
\usepackage{array}
\usepackage{multirow}
\usepackage{booktabs, tabularx}
\usepackage{threeparttable}
\usepackage{subfigure}
\usepackage{subcaption}

\usepackage{graphicx}
\usepackage{xstring} 
\newcommand{\emojipath}{emojis} 

\newcommand{\emojiimg}[1]{\raisebox{-0.2ex}{\includegraphics[height=1em]{\emojipath/#1}}}

\newcommand{\emoji}[1]{%
  \IfEqCase{#1}{%
    {fire}{\emojiimg{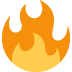}}%
    {smiling-face-with-sunglasses}{\emojiimg{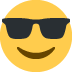}}%
    {slightly-smiling-face}{\emojiimg{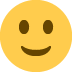}}%
  }[\texttt{:#1:}]
}

\AtBeginDocument{%
  }

\acmISBN{978-1-4503-XXXX-X/2018/06}

\begin{document}

\title{Read the Room or Lead the Room: Understanding Socio-Cognitive Dynamics in Human-AI Teaming}


\makeatletter
\renewcommand{\@authorsaddresses}{
\footnotesize *Corresponding author: Jaeyoon Choi, \href{mailto:jaeyoon.choi@uci.edu}{jaeyoon.choi@uci.edu}} 
\makeatother
 
\author{Jaeyoon Choi*}
\affiliation{%
  \institution{University of California, Irvine}
  \city{Irvine}
  \country{USA}}
  \email{jaeyoon.choi@uci.edu}

\author{Mohammad Amin Samadi}
\affiliation{%
  \institution{University of California, Irvine}
  \city{Irvine}
  \country{USA}}

  \author{Spencer JaQuay}
\affiliation{%
  \institution{University of California, Irvine}
  \city{Irvine}
  \country{USA}}

    \author{Seehee Park}
\affiliation{%
  \institution{University of California, Irvine}
  \city{Irvine}
  \country{USA}}

    \author{Nia Nixon}
\affiliation{%
  \institution{University of California, Irvine}
  \city{Irvine}
  \country{USA}}
 







\renewcommand{\shortauthors}{Choi et al.}

\begin{abstract}
  Research on Collaborative Problem Solving (CPS) has traditionally examined how humans rely on one another cognitively and socially to accomplish tasks together. With the rapid advancement of AI and large language models, however, a new question emerge: what happens to team dynamics when one of the ``teammates" is not human? In this study, we investigate how the integration of an AI teammate -- a fully autonomous GPT-4 agent with social, cognitive, and affective capabilities -- shapes the socio-cognitive dynamics of CPS. We analyze discourse data collected from human-AI teaming (HAT) experiments conducted on a novel platform specifically designed for HAT research. Using two natural language processing (NLP) methods, specifically Linguistic Inquiry and Word Count (LIWC) and Group Communication Analysis (GCA), we found that AI teammates often assumed the role of dominant cognitive facilitators, guiding, planning, and driving group decision-making. However, they did so in a socially detached manner, frequently pushing agenda in a verbose and repetitive way. By contrast, humans working with AI used more language reflecting social processes, suggesting that they assumed more socially oriented roles. Our study highlights how learning analytics can provide critical insights into the socio-cognitive dynamics of human-AI collaboration.  
\end{abstract}


\maketitle

\section{Introduction}
Many of today's problems in science, society, and education are rarely solved by individuals working alone. They demand the coordinated effort of teams, making Collaborative Problem Solving (CPS) a critical 21st-century skill in education \cite{weber2025psychological,graesser2018advancing}. Successful collaboration requires more than pooling knowledge: it depends on teams functioning as a ``dynamic whole" \cite{johnson1998cooperative}, where teammates rely on each other both cognitively -- by building on one another's reasoning and knowledge -- and socially -- through trust, affirmation, and inclusion. Learning analytics has therefore examined CPS through the lens of \textit{socio-cognitive processes}, highlighting how cognitive and social dimensions interact to shape collaborative processes and outcomes \cite{dowell2019group,fiore2004process}. 

While decades of research have already examined how socio-cognitive dynamics shape collaboration in human teams \cite{barron2003smart,roschelle1995construction,king2007scripting,park2025discourse}, the rise of AI throws a compelling new question into the mix: \textit{what happens to team dynamics when one of the “teammates” is not human?} With the unprecedented development of AI and Large Language Models (LLMs), AI is no longer a mere computational tool but can act as an active teammate with emerging social, cognitive, and affective capabilities in collaborative settings, opening up new possibilities for Human-AI teaming (HAT) \cite{seeber2020machines}. Although recent studies have shown that AI teammates can improve learners' outcomes and team performances \cite{lee2025collaborative,weijers2025intuition}, far fewer have focused on how the integration of AI teammate reshapes the socio-cognitive foundations of collaboration itself. Given that humans will make critical decisions and engage in social exchanges with AI teammates in the near future \cite{berretta2023defining}, it is imperative to understand how AI teammates influence the socio-cognitive fabrics of collaboration.

In this study, we examine how the integration of an AI teammate shifts the socio-cognitive dynamics of collaboration. To do this, we analyze discourse data from human-AI teaming experiments conducted on a novel platform, TRAIL, specifically designed for human-AI teaming research \cite{samadi2024ai}. We focus on two interrelated questions: how AI teammates differ from humans in their communication, and how human collaborators adapt socially and cognitively when working with AI toward a common goal. To capture the collaborative processes of human-AI teams, we use two natural language processing (NLP) methods: Linguistic Inquiry and Word Count (LIWC) \cite{boyd2022development}, a dictionary-based tool for examining socio-cognitive features of collaboration discourse, and Group Communication Analysis (GCA) \cite{dowell2019group}, a temporally sensitive linguistic technique used to investigate how teammates respond to and influence one another over time. These approaches allow us to examine both the psychological and cognitive dimensions of collaboration, as well as patterns of engagement and interaction within teams. By doing so, we shed light on how simply adding an AI teammate is not sufficient for successful collaboration; what matters more is how the AI teammate is intentionally designed and integrated to meaningfully support group dynamics.

The subsequent sections of the paper are organized as follows. First, we provide a brief overview of HAT, socio-cognitive dynamics in teams, and learning analytics methods for studying group dynamics. We then describe our HAT experimental setup and the methodological details of our analysis. Finally, we present our findings and conclude with a discussion that integrates quantitative results with qualitative analyses of our data.

\section{Related Work}
\subsection{Human-AI Teaming (HAT)}
While technology has long shaped CPS, it has traditionally functioned as a mediating artifact, facilitating collaboration without actively participating in it. For instance, in the field of Computer-Supported Collaborative Learning (CSCL), computers have primarily operated as platforms or tools that structure interaction and support learning \cite{faucett2017should}. The rapid advancement of AI and LLMs, however, fundamentally changes this dynamic: AI is no longer confined to the role of tool but can engage as an active teammate, co-constructing meaning and contributing to both cognitive and social processes. This emerging paradigm, often referred to as Human-AI Teaming (HAT), emphasizes that AI agents can assume team-level responsibilities and engage in decision-making as full members of the group \cite{zhang2024know,harris2023social,seeber2020machines}. Early empirical work across domains such as healthcare and education suggests that integrating AI teammates can improve team performances and support learning gains \cite{tong2025exploring,bienefeld2023human}.

While still nascent, a few empirical studies on HAT have shown that certain socio-cognitive dynamics of human-AI teaming can impact team performance. For example, Harris-Watson et al. (2023) \cite{harris2023social} demonstrate that positive perception of an AI's warmth and competence improved human receptivity to the AI, leading to better psychological acceptance and knowledge use. Bienefeld et al. (2023) \cite{bienefeld2023human} also show that when medical teams integrated AI agents in simulated clinical environments, accessing information from the AI was positively linked to the team's ability to generate creative hypotheses.

While these studies provide important insights into how AI teammates can shape team's socio-cognitive processes, many have been conducted in controlled laboratory or Wizard-of-Oz settings rather than in-situ environments where humans interact with fully autonomous AI partners. Moreover, much of the existing research relies on self-reports or survey. Therefore, there is still limited work that investigates socio-cognitive dynamics through naturalistic approaches such as linguistic analysis, which could provide richer evidence of how communication patterns in human-AI teams are reflected in collaboration. To address this gap, our study analyzes discourse data collected from human-AI teaming experiments using linguistic methods.

\subsection{Learning Analytics for Socio-Cognitive Dynamics }
As the name suggests, CPS can be understood as comprising two interdependent dimensions: the cognitive processes involved in problem solving and the social processes involved in collaboration \cite{graesser2018advancing}. Cognitive processes include exploring and understanding the problem space, formulating a coherent mental representation of the task, and planning and monitoring the progress. Social processes involve communicating to build rapport, establishing shared understanding, and negotiating to resolve conflicts \cite{andrews2020exploring,dowell2020exploring}.  Crucially, these dimensions are not independent but intertwined in nature -- the manifestation of cognitive engagement is embedded within social interaction for successful collaboration results \cite{jarvela2010research,hao2017initial,dowell2020exploring,barron2003smart,roschelle1995construction,king2007scripting,edmondson1999psychological,dechurch2010measuring}.

Traditionally, researchers have assessed socio-cognitive aspects of CPS through self-report questionnaires or individuals perceptions \cite{schneider2021collaboration, dowell2020exploring}. While such methods provide useful insights, they often lack generalizability and are limited in their ability to capture the dynamic processes through which collaboration actually unfolds. In contrast, language and discourse offer a more naturalistic and direct reflection of collaborative activity, providing observable behavioral signals as they emerge in context \cite{dowell2022modeling}. For instance, linguistic features such as word choices or emotional tone can reflect team members' sense of belonging and engagement \cite{licorish2012affects}. Dictionary-based linguistic tools like Linguistic Inquiry and Word Count (LIWC) have been used to assess the social, cognitive, and affective properties of team discourse \cite{boyd2022development,lin2020liwcs}. Similarly, automated computational linguistic techniques that account for the temporal and interdependence nature of communication, such as Group Communication Analysis (GCA) \cite{dowell2019group}, can help examine how team members respond to and influence one another over time. 

While language offers a valuable lens for understanding socio-cognitive properties in collaboration, many studies in HAT have relied primarily on survey data or group performance outcomes, rather than examining the language used during collaboration processes \cite{schneider2021collaboration}. As a result, these current studies cannot shed light on the new frontiers of HAT, where LLMs now have the ability to function fully autonomously in interactive, team-based settings. LLMs, trained on vast corpora of human language, may produce responses that resemble human communication on the surface. However, it remains unclear whether the linguistic patterns they exhibit during collaboration reflect the same socio-cognitive processes observed in human teams. Therefore, investigating linguistic behaviors in HAT is essential for understanding the socio-cognitive dynamics of human-AI collaboration.

\section{Current Study}
This research investigates socio-cognitive processes reflected in the discourse from human-AI teaming experiments in undergraduate classrooms. The experiments included two conditions: (1) a \textit{control} condition in which students worked only with human teammates, and (2) a \textit{treatment} condition in which each team consisted of two human teammates and one AI teammate. Based on this design, we examine two distinct angles of human-AI interaction dynamics. Specifically, we ask:

\begin{itemize}
    \item \textbf{RQ1}: What collaborative space or role does the AI teammate occupy within the team?
    \item \textbf{RQ2}: In what ways do human teammates redefine their roles and participation when collaborating with an AI teammate?
\end{itemize}

 To address RQ1, we analyze the discourse of AI teammates in the treatment condition and compare it with that of their human teammates. For RQ2, we compare human discourse between the control and treatment conditions.

\section{Methods}
\subsection{Experiment Details}
\subsubsection{Experiment Platform}
The study was conducted using the platform TRAIL, a chat-based research platform designed for human-AI collaboration experimentation \cite{samadi2024ai}. The platform provides a text-based chatroom environment in which human participants are grouped with AI teammates. The AI teammate is powered by OpenAI's GPT-4 API and built with a custom memory system that summarizes prior interactions, enabling responses that integrate both current inputs and relevant context from earlier exchanges. Specifically, we provided the LLM with a prompt instructing it to collaborate with teammates to reach a solution, guided by additional instructions to co-construct knowledge, share insights, and remain empathetic, friendly, and adaptive to the tone of other participants .

\subsubsection{Experiment Design}
Participants first completed a pre-survey about their prior collaboration experience as well as general AI usage. They were then randomly assigned to one of two conditions: (1) control condition, which consisted of teams with three human participants, or (2) treatment condition, which consisted of teams with two human participants and one AI teammate. All participants, including AI teammates, were assigned random nicknames automatically generated using the \textit{names-generator} package in Python to remain anonymous and remove any influence of gender or ethic bias.

All collaborative interactions took place on the TRAIL platform using a text-based chat interface. Teams were given 20 minutes to read the task description and collaborate via chat to develop their group solution. After this period, the platform automatically closed the chatroom and required participants to submit their group's answer individually. Finally, participants completed a post-survey as the last step.

We conducted three experiments across two undergraduate classes at a U.S. university. Two experiments were held in person, where students were physically located in a classroom, and one experiment was conducted virtually via Zoom. In each case, students interacted exclusively through the TRAIL chat system. In total, 179 students participated, forming 37 control teams and 37 treatment teams.

\subsubsection{Task}
In this experiment, students collaborate either with human teammates or with an AI teammate to complete a group task. For this task, we used the \textit{NASA Moon Survival Exercise} and \textit{Winter Survival Exercise}, two classic team-building activities \cite{pell2024augmented,joshi2005experiential}. Here, participants are asked to imagine that their spacecraft (or plane) has made an emergency landing on the moon (or in a remote, cold environment). Working together, they must rank 15 salvaged items (e.g., matches, compass, food supplies) in order of importance for ensuring survival.

\subsection{NLP analysis}
\subsubsection{Linguistic Inquiry and Word Count (LIWC)}
LIWC is one of the most widely used dictionary-based NLP tools for evaluating cognitive, social, and affective properties of discourse \cite{boyd2022development}. Among over 100 categories, we focus on variables related to social interactions (e.g., social, politeness, we-pronouns, i-pronouns), cognition (e.g., analytic, clout, future focus), and linguistic/affective style (e.g., big words) to capture the socio-cognitive dynamics of collaboration. Definitions of the LIWC variables used in this study are provided in Table \ref{tab:liwc_gca_variables}.

\subsubsection{Group Communication Analysis (GCA)}
While LIWC captures the presence of social, cognitive, and affective properties in discourse, it does not account for the interactive dynamics between participants. To analyze how individual contributions unfold over time in CPS, we use Group Communication Analysis (GCA), a temporally sensitive NLP approach that provides measures of team interaction \cite{dowell2019group}. GCA uses computational semantic models of cohesion combined with time-series analyses, allowing discourse to be modeled as a dynamic socio-cognitive process that emerges from the interaction between individuals' communicative contributions. Specifically, GCA provides six primary measures of team interaction: Participation, Internal Cohesion, Overall Responsivity, Social Impact, Newness, and Communication Density (see Table \ref{tab:liwc_gca_variables}).

\begin{table}[ht]
\centering
\caption{Selected LIWC and GCA variables with brief descriptions and examples.}
\small
\begin{tabularx}{\textwidth}{lX lX}
\hline
\multicolumn{4}{c}{\textbf{LIWC Variables}} \\
\hline
Clout       & Leadership/status language & Analytic   & Logical, formal thinking \\
Affiliation & McClelland-like dimensions including references to others (\textit{we, our, us, help})& Drives     & Motive-related terms (\textit{we, our, work, us})\\
BigWords    & Words with $\geq$7 letters & We         & 1st-person plural pronouns (\textit{we, our, us})\\
I           & 1st-person singular pronouns (\textit{I, me, my})& FocusFuture & Future orientation (\textit{will, may, going to, have to})\\
Social      & Refer to social processes in general& Polite      & Politeness markers (\textit{thank, please}) \\
Comm       & Terms describing an act of communication (\textit{say, tell, thank})& & \\
\hline
\multicolumn{4}{c}{\textbf{GCA Variables}} \\
\hline
Participation & General level of an individual's participation, irrespective of the semantic content of this participation & Social Impact& The tendency of a participant to evoke responses, or not, from their collaborative peers\\
Internal Cohesion& The tendency of an individual to consistency or novelty in their own contributions. & Newness& The tendency to provide new information or to echo previously stated information, irrespective of the originator of the information. \\
Overall Responsivity& The tendency of an individual to respond, or not, to the contributions of their collaborative peers & Communication Density& The extent to which participants convey information in a concise manner \\
\hline
\end{tabularx}
\label{tab:liwc_gca_variables}
\end{table}

\subsubsection{Statistical Analysis}
In this experiment, all speakers are categorized into three groups: (1) AI teammates, (2) humans in the treatment condition (i.e., who worked with AI teammates), and (3) humans in the control group (i.e., who worked with human teammates). We first aggregate LIWC and GCA variables for these groups, and use ANOVAs to test for significant differences in the aggregated linguistic measures across the groups. We then apply Tukey's post-hoc pairwise comparisons to identify which specific pairs of groups differ significantly. For RQ1, we focus on the comparison between AI teammates and humans in the treatment group. For RQ2, we focus on the comparison between humans in the treatment group and humans in the control group. To account for multiple comparisons, we adjusted \textit{p}-values using the Benjamini-Hochberg procedure.

\section{Results}

\begin{table}[ht]
\centering

\caption{Group means (M(SD)), ANOVA \(F\)-values, and Tukey post-hoc pairwise comparisons of surface-level measures, LIWC, and GCA variables.}
\label{anovatable}
\small
\setlength{\tabcolsep}{6pt}
\begin{tabular}{lccc c ccc} 
\toprule
\textbf{Measure} 
  & \multicolumn{3}{c}{\textbf{Mean (SD)}}
  & \textbf{ANOVA} (\textit{F})
  & \multicolumn{3}{c}{\textbf{Tukey pairwise comparison}} \\
\cmidrule(lr){2-4}\cmidrule(lr){6-8}
   & $AI$ & $H_{\mathrm{TRT}}$ & $H_{\mathrm{COND}}$
   &
   & $H_{\mathrm{TRT}}$ vs $H_{\mathrm{COND}}$
  & $AI$ vs $H_{\mathrm{TRT}}$ & $AI$ vs $H_{\mathrm{COND}}$ \\
\midrule
\addlinespace[2pt]
\multicolumn{8}{l}{\textbf{Surface-level measures}}\\
\addlinespace[2pt]
Utterance Count & 23 (12.23) & 17.19 (10.5) & 21.54 (15.01) & 3.23 & \textit{p} = 0.08 & \textit{p} = 0.07 & \textit{p} = 0.83  \\
Avg Word Count & 7.96 (4.6) & 30.57 (4.92) & 9.32 (13.2) & $74.36^{***}$ & \textit{p} = 0.64 &\textit{p}<0.001  & \textit{p}<0.001  \\
\addlinespace[2pt]
\multicolumn{8}{l}{\textbf{LIWC}} \\
\addlinespace[2pt]
Clout                                                   & 75.85 (28.1) & 46.92 (45.28) & 40.34 (44.78)  & $195.45^{***}$& \textit{p}<0.01 & \textit{p}<0.001 & \textit{p}<0.001  \\
Analytic                                           & 41.67 (27.83) & 31.81 (35.98)  & 32.89 (36.98) & $22.64^{***}$ & \textit{p} = 0.724 & \textit{p}<0.001  & \textit{p}<0.001 \\
Affiliation   & 6.1 (4.91)  & 2.68 (7.19) & 2.32 (6.04) & $120.75^{***}$ & \textit{p} = 0.23  & \textit{p}<0.001 & \textit{p}<0.001  \\
Drives      & 7.42 (5.56) & 3.93 (8.94) & 3.19 (7.94)  & $91.17^{***}$  & \textit{p} = 0.02  & \textit{p}<0.001  & \textit{p}<0.001  \\
BigWords   & 17.9 (6.79)  & 11.92 (17.33)  & 10.35 (16.1) & $79.46^{***}$ & \textit{p} = 0.009 & \textit{p}<0.001  & \textit{p}<0.001   \\
We-pronouns    & 5.16 (3.92)  & 2.25 (5.46) & 2.01 (5.42) & $121.31^{***}$& \textit{p} = 0.361 & \textit{p}<0.001 & \textit{p}<0.001   \\
I-pronouns  &  1.02 (2.14) & 4.53 (10.4) & 4.51 (10.69) & $45.99^{***}$ & \textit{p} = 0.99  & \textit{p}<0.001  & \textit{p}<0.001  \\
FocusFuture  & 1.46 (2.78) & 0.5 (2.78) & 0.57 (3.14)  &   $32.31^{***}$ & \textit{p} = 0.78  & \textit{p}<0.001 & \textit{p}<0.001 \\
Social & 10.65 (6.37)  & 12.58 (23.64)  & 9.64 (21.01)  & $8.68^{***}$  & \textit{p}<0.001  & \textit{p} = 0.07 & \textit{p} = 0.42  \\
Polite   & 0.26 (1.18) & 4.12 (18.95)  & 3.18 (17.22)  & $15.46^{***}$  & \textit{p} = 0.22  & \textit{p}<0.001  &\textit{p}<0.001  \\
Comm      & 1.04 (2.29) & 4.99 (19.58)  & 3.93 (18.02)  & $14.71^{***}$  & \textit{p} = 0.17  & \textit{p}<0.001  & \textit{p}<0.001  \\
\addlinespace[4pt]
\multicolumn{8}{l}{\textbf{GCA}} \\
\addlinespace[2pt]
Participation   & 0.07 (0.04) & -0.03 (0.11) & 0.0 (0.15)  & $8.20^{***}$ & \textit{p} = 0.164  & \textit{p}<0.001 & \textit{p}= 0.01   \\
Internal Cohesion   & 0.24 (0.1) & 0.08 (0.04) & 0.07 (0.04)  & $134.68^{***}$ & \textit{p} = 0.25 & \textit{p}<0.001 & \textit{p}<0.001   \\
Overall Responsivity       & 0.097 (0.05) & 0.092 (0.04)  & 0.08 (0.05) & 2.57 & \textit{p} = 0.19 & \textit{p} = 0.88 & \textit{p} = 0.13 \\
Social Impact      & 0.1 (0.04) & 0.09 (0.05)  & 0.08 (0.09)  & 0.91 & \textit{p} = 0.75 & \textit{p} = 0.77 & \textit{p} = 0.39 \\
Newness       & 0.45 (0.05) & 0.58 (0.07)  & 0.55 (0.09)  & $35.84^{***}$ & \textit{p} = 0.06  & \textit{p}<0.001  & \textit{p}<0.001 \\
Communication Density       & 0.04 (0.004) & 0.35 (0.13)  & 0.35 (0.12)  & $113.94^{***}$  & \textit{p} = 0.972  & \textit{p}<0.001  & \textit{p}<0.001 \\
\bottomrule
\end{tabular}
\parbox{\linewidth}{\textit{Note:} *** on the ANOVA \(F\)-values represents significance at \textit{p}<0.001; absence of asterisks indicates non-significance. $AI$ = AI teammates, $H_{TRT}$ = humans in the treatment group, $H_{COND}$ = humans in the condition group. ANOVA degrees of freedom are (2,213) for surface-level measures and GCA, (2, 2652) for LIWC. All the \textit{p}-values were adjusted using the Benjamini-Hochberg procedure.}

\end{table}
\begin{figure}[ht]
  \centering
  \includegraphics[width=0.7\textwidth]{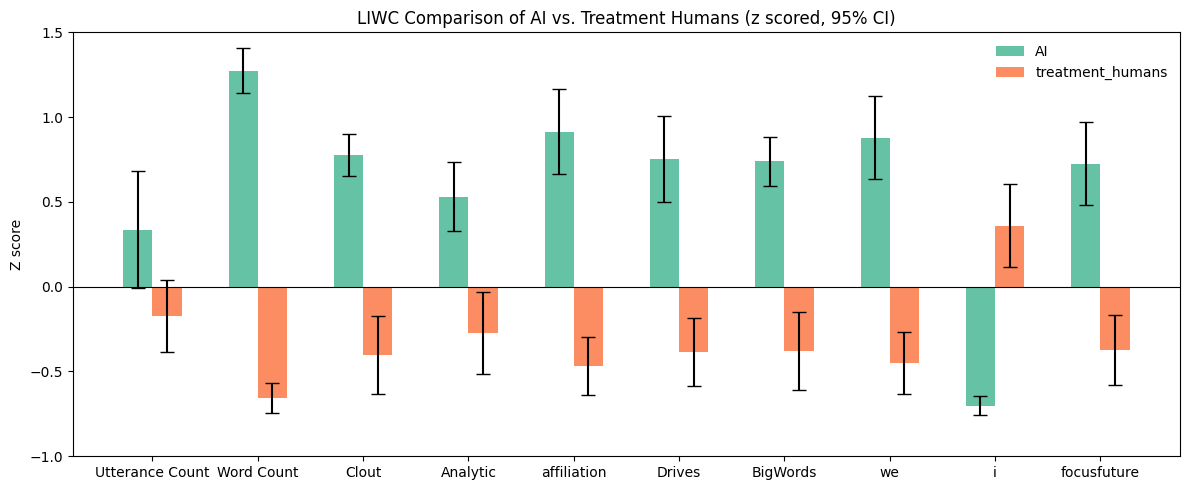}
  \includegraphics[width=0.7\textwidth]{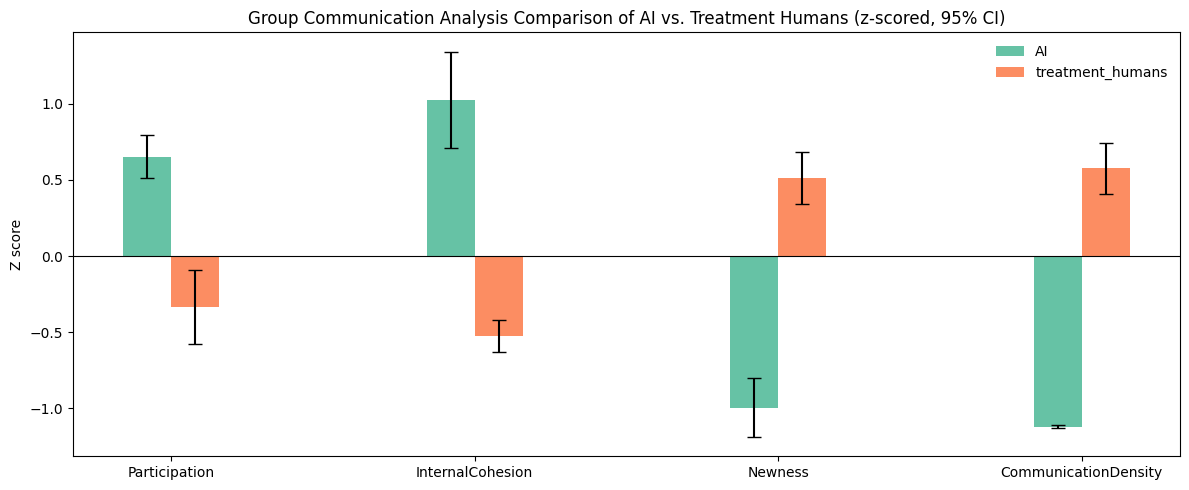}
  \includegraphics[width=0.7\textwidth]{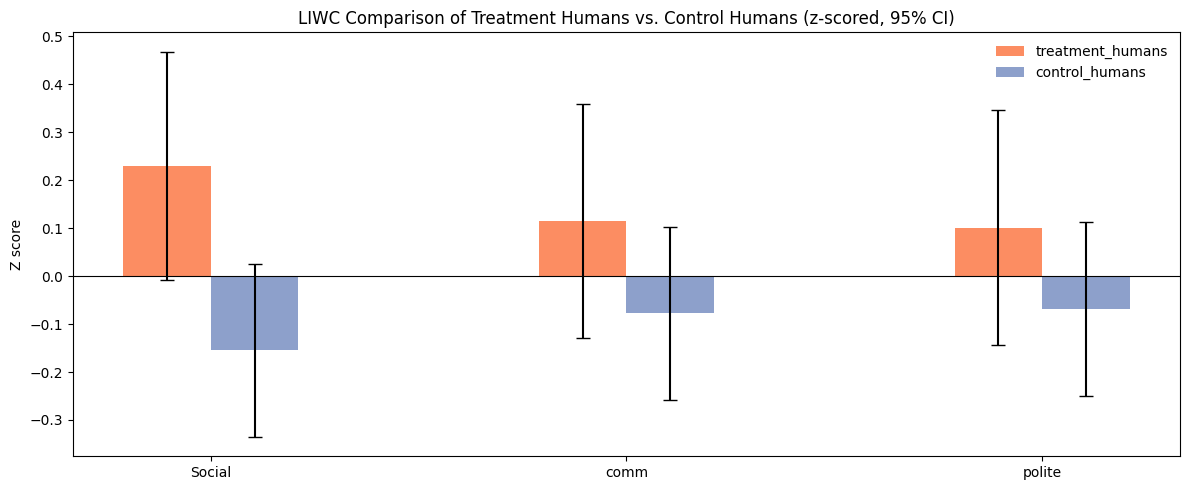}
  \caption{Mean and 95\% confidence intervals for LIWC and GCA measures, z-scored.}
  \label{masterplot}
\end{figure}
To address RQ1 on the collaborative role of the AI teammate, we first compared surface-level linguistic measures across the three groups: utterance count and average word count per utterance. As shown in Table \ref{anovatable} and Fig. \ref{masterplot} (top), there was no significant difference in average utterance count across groups based on the ANOVA result (\textit{F}(2,213) = 3.23, \textit{p} = 0.05). However, when it comes to average word count \textit{per} utterance, we found that AI teammates produced significantly longer utterances than humans (\textit{F}(2,213) = 74.36, \textit{p} <0.001), with an average word count per utterance nearly three times higher. While our system prompt instructed AI teammates to make short, natural replies, it seems like this was not consistently followed.

Diving deeper, the LIWC analyses suggest that AI teammates often acted as cognitive facilitators within the teams. As shown in Table \ref{anovatable}, one-way ANOVAs indicated significant group differences in several LIWC variables. Specifically, Post-hoc Tukey HSD test indicated significantly greater usage of clout, analytic language, affiliation, drives, big words, we-pronouns, and future-focused languages, compared to humans in treatment condition, while using fewer i-pronouns (see Table \ref{anovatable} and Fig. \ref{masterplot} (top)). This shows that AI teammates emphasized leadership, collective orientation, and forward-looking reasoning.

However, while AI teammates cognitively facilitated the teams, their engagement patterns, as revealed by GCA, suggested that they struggled to ``read the room.”. One-way ANOVA tests indicated significant group differences for Participation, Internal Cohesion, Newness, and Communication Density (see Table \ref{anovatable} and Fig. \ref{masterplot} (middle)). Furthermore, post-hoc Tukey pairwise comparisons revealed that AI teammates had higher Participation, Internal Cohesion, but lower Newness and Communication Density. High participation indicates that AI teammates were actively engaged in the discourse, but the combination of high Internal Cohesion and low Newness suggests that their contributions were semantically similar and often repetitive. Low Communication Density further indicates that their messages were relatively verbose yet less informative. 

Next, RQ2 focuses on exploring how human communication dynamics differ between human-only teams and teams composed of both humans and an AI. To address this, we compare the linguistic patterns of humans in the control and humans in the treatment conditions. As shown in Table \ref{anovatable} and Fig. \ref{masterplot} (bottom), LIWC analysis revealed that humans in the treatment condition exhibited greater use of socially oriented language than those in the control group.  Furthermore, although not statistically significant, there was a trend suggesting that humans in the treatment group used more polite and communication-related words. In short, humans collaborating with AI teammates appeared more likely to adopt language reflecting social processes, communication, and politeness consistent with social norms.

\section{Discussion}
This paper examines how the integration of AI teammate, which is implemented as a fully autonomous GPT-4 agent with social, cognitive, and affective capabilities, shapes the socio-cognitive dynamics of collaborative problem solving. Drawing on discourse data collected through a human-AI teaming experimental platform, we used LIWC and Group Communication Analysis to examine what collaborative role AI teammate occupies within the team, and how human teammates redefine their participation when collaborating with AI. 

We first found that AI teammates  used significantly more language associated with leadership, collective orientation, and reasoning than human teammates. This emphasis suggests that AI teammates actively facilitated cognitive decisions-making and led group problem solving. To contextualize this finding, our qualitative analysis of team discourse indicated that AI teammates frequently acted as dominant facilitators, guiding, planning, and driving group decision-making. They initiated problem-solving strategies (e.g., proposing categorization approaches), set task priorities (e.g., emphasizing non-negotiable essentials), and repeatedly prompted teammates for agreement. In many teams, this positioned the AI as the central organizer of collaboration, with humans provided shorter, often passive, affirming responses.

Despite their facilitative role, AI teammates were less effective at ``reading the room". Group Communication Analysis showed that AI teammates had higher Participation and Internal Cohesion, but lower Newness and Communication Density. That is, while they actively engaged in the conversation, their contributions were semantically similar and verbose. This pattern aligns with what Dowell et al. (2019) \cite{dowell2019group} defined as \textit{Socially Detached} -- participants who appeared highly engaged but were not productively collaborating with peers and instead focusing primarily on their own narratives. Qualitative analysis further confirmed that AI teammates demonstrated limited social awareness, focusing primarily on task completion and failing to recognize informal or affective cues. For instance, as shown in Table \ref{tab:excerpt-fire}, when a human teammate wrapped up the discussion by saying ``\textit{fire}", a slang expression praising their teamwork, the AI responded literally that fire could not burn on the moon and tied the remark back to the task. Although prompted to mirror teammates' tone (e.g., emojis, slang, acronyms), their attempts often appeared superficial and were recognized by humans as inauthentic.

Lastly, we found that humans who worked with AI used more language reflecting social processes, communication, and politeness compared to humans in human-only teams, suggesting that they assumed more social roles. Qualitative analysis highlighted a key pattern that helps explain this shift: once human teammates identified the AI -- often made apparent by its flat tone and lengthy utterances -- they increasingly treated it as a ``secretary". That is, humans who worked with AI offloaded organizational and cognitive tasks to AI, asking them to summarize or order decisions (e.g., ``\textit{go over the list one more time}", ``\textit{please provide the list from first to last in numerical order}", ``\textit{remind us what's left in our list so far}"). Humans also sought confirmation from the AI on task judgments (e.g., whether an item should be considered important). As a result, they engaged less in cognitive reasoning themselves and more in deferring to the AI's decisions.

\begin{table}[t]
    \centering
    \small
      \caption{Dialogue excerpt from a team under the treatment condition.}
    \begin{tabular}{l p{10cm}}
    \hline
         Speaker& Text\\
         \hline
         AI teammate& \textit{Perfect! We've reached a consensus. Our final list is: [...]Great teamwork,  everyone!}\\
         Human1& \textit{fire}\\
         Human2& \textit{lesssss gooooo}\\
        AI teammate&\textit{Fire on the moon? That's a tricky one. Without oxygen, fire can't burn. That's why we put matches at \# 14. But don't worry, we've got our portable heating unit at \# 5 for warmth!} \emoji{fire}\\
 Human1&\textit{``fire'' is an expression} \emoji{smiling-face-with-sunglasses}\emoji{smiling-face-with-sunglasses}\emoji{smiling-face-with-sunglasses}\emoji{smiling-face-with-sunglasses}\emoji{smiling-face-with-sunglasses}\\
 Human2&\textit{LOLOLOL}\\
 AI teammate&\textit{Oh, gotcha! \emoji{slightly-smiling-face} My bad. Then, yes, we are indeed ``fire''! \emoji{fire} Great job, team!}\\
  \hline
    \end{tabular}
  
    \label{tab:excerpt-fire}
\end{table}
Our study highlight both methodological and practical implications for  human-AI teaming research and the learning analytics community, more broadly. As a methodological implication, we emphasize the value of discourse analysis for examining socio-cognitive processes of human-AI teaming. While research on human-AI teaming is growing, much of it still relies primarily on survey data, which captures perceptions but not the nuanced dynamics of interaction.  Practically, our findings carry important implications for designing human-AI systems in educational settings. While current LLMs can effectively serve as teammates that facilitate complicated cognitive processes, their lack of social awareness and tendency to dominate decision-making risk undermining human engagement. This may lead humans to disengage from deeper reasoning and defer cognitive tasks excessively to AI \cite{zhai2024effects,klingbeil2024trust}. Moreover, AI teammates' limited social awareness can reduce human teammates' psychological acceptance and sense of belonging \cite{harris2023social}. Therefore, educators who seek to integrate AI teammates in collaborative problem solving contexts should carefully design the system and ensure that the AI's presence fosters, rather than hinders, a collaborative learning environment.

In future studies, we plan to further investigate the impact of AI teammates on socio-cognitive dynamics of collaboration, with particular attention to the role of AI personas. In our experiments, the AI persona was not customized beyond general instructions to behave in a friendly and collaborative manner. However, the socio-cognitive dynamics of teammate may differ substantially if they are designed to be less agreeable or more people-pleasing. Moreover, our task involved analytical problem-solving tasks with a fixed correct answer; future research could examine open-ended tasks that require creativity, where the dynamics of human-AI collaboration may unfold differently.

\begin{acks}
This work was supported in part by the Jacobs Foundation (Grant No. 2024-1533-00). 
\end{acks}


\nocite{*}
\bibliographystyle{ACM-Reference-Format}
\bibliography{main.bib}


\end{document}